\documentclass[aps,twocolumn,superscriptaddress,showpacs,floatfix]{revtex4}
\usepackage{graphicx}
\begin{document}

\title{Distributions of inherent structure energies during aging}

\author{Ivan Saika-Voivod}

\affiliation{Dipartimento di Fisica, Istituto Nazionale per la
Fisica della Materia, and INFM Center for Statistical Mechanics and
Complexity, Universit\`a di Roma La Sapienza, Piazzale Aldo
Moro~2, I-00185, Roma, Italy}

\author{Francesco Sciortino}
\affiliation{Dipartimento di Fisica, Istituto Nazionale per la
Fisica della Materia, and INFM Center for Statistical Mechanics and
Complexity, Universit\`a di Roma La Sapienza, Piazzale Aldo
Moro~2, I-00185, Roma, Italy}

\date{\today}

\begin{abstract}
We perform extensive simulations of a binary
mixture Lennard-Jones system subjected to a temperature jump in order to study
the time evolution of fluctuations during aging.
Analyzing data from 1500 different aging realizations, we
calculate distributions of inherent structure energies for different 
aging times and contrast them with equilibrium.  We
find that the distributions initially become narrower
and then widen as the system equilibrates.  
For deep quenches, fluctuations in the glassy system differ
significantly from those observed in equilibrium.
Simulation results are partially captured by theoretical predictions 
only when the final temperature is higher than the mode coupling temperature.
\end{abstract}


\maketitle
When a liquid is rapidly brought out of equilibrium,  for example by an abrupt 
change in temperature $T$, it starts to age. The material properties change 
with the aging time $t_{\rm w}$
in the attempt to recover equilibrium at the 
new bath temperature $T_{\rm b}$.  This aging process can be visualized as a 
path in configurational space, that starts from a typical equilibrium 
configuration before the quench and which, at long times, should 
converge to an equilibrium configuration characteristic of $T_{\rm b}$.   
When $T_{\rm b}$ is below the 
glass transition temperature, the aging dynamics never stops 
in the experimentally accessible time window, but becomes 
slower and slower on increasing $t_{\rm w}$.

In recent years, several studies have revisited the physics of 
aging~\cite{tool46,brawer,scherer86,mckenna89}, 
attempting to associate the aging dynamics with a progressive decrease 
of the fictive (or effective) temperature 
$T_{\rm f}$~\cite{cug97,fra00,nie01,fsfdt,mos02}. 
In these studies, the configuration of the aging system 
at time $t_{\rm w}$ is associated with a typical  configuration 
explored by the 
equilibrium liquid at $T_{\rm f}$. In this case, 
the aging dynamics can be modeled as a progressive thermalization of the 
system, quantified by the progressive decrease of $T_{\rm f}$. 

A convenient framework for  analyzing the thermodynamics of supercooled 
liquids and aging dynamics in numerical simulations is  provided by 
the inherent structure (IS) formalism~\cite{sti82}.
In this framework, each  configuration 
is associated with its closest local minimum, named IS, on the potential 
energy surface (PES) of the liquid.  A formal 
thermodynamic description of the system can be developed with a proper 
modeling of  the statistical properties of the PES, i.e., the distribution 
of the energy minima $e_{\rm IS}$ and of their basins of 
attraction~\cite{sti82,bacini}. 
It has been shown that in equilibrium supercooled states, the probability 
density $P(e_{\rm IS},T)$
of exploring a minimum of depth $e_{\rm IS}$ 
is approximately 
Gaussian, with a $T$-independent variance and with an average  
$\bar{e}_{\rm IS}$ which decreases monotonically 
with $T$~\cite{sci99,heu00,sri01}.  
In the IS 
approach, the aging dynamics can be visualized as the progressive exploration 
of deeper and deeper minima, and a connection between aging 
and equilibrium configurations can be made by comparing 
properties of the explored basins.

Experimental studies of glass forming materials~\cite{kovacs,mckenna89} 
and recent numerical  simulations~\cite{mossa03} have called attention to 
the fact that aging systems,  especially for deep quenches, may not be 
associated with a single $T_{\rm f}$, i.e., can not be uniquely associated 
with a liquid  configuration, 
calling for a further development of the present theories.  
In this Letter, we bring  the level of comparison between equilibrium and
aging to a much more detailed level, by studying
fluctuations around the average properties. We show that the fluctuations
frozen in the glass depend on the thermal history and are significantly 
different from the fluctuations characteristic of the liquid state. 


We perform molecular dynamics (MD) simulations of $N=1000$ particles
in a cubic box of length $L=9.4$,
for the well characterized Kob-Andersen binary-mixture 
Lennard-Jones  (LJ) model~\cite{kob95},  
for which the  mode coupling transition has been estimated at 
$T_{\rm x}=0.435$.
We employ the Nos\'e-Hoover thermostat with parameters chosen so that 
the kinetic energy thermalizes 
in a time between $300$ and $1000$ time steps.  Time is reported here in
number of MD time steps, each of which is $0.01 t_0$~\cite{units}. 
We calculate the probability density $P(e_{\rm IS},t_{\rm w})$ of 
finding the liquid at a given  time $t_{\rm w}$ in a basin of energy 
$e_{\rm IS}$,  as the system responds to a sudden lowering in $T$.  
To this end, we perform an ensemble of 
$1500$ independent MD runs, each for $10^6$ steps. 
The temperature is switched from the initial equilibrium temperature
$T_{\rm i}$ to the new $T_{\rm b}$ at $t_{\rm w}=0$.
For each MD run we select configurations at $35$ different  $t_{\rm w}$, 
logarithmically spaced.
We perform
a conjugate-gradient quench on each configuration to obtain the associated IS.
Diagonalization of the Hessian matrix  evaluated at the IS provides the
eigenfrequencies $\{ \omega_i \}$.
Thus for each studied 
$t_{\rm w}$ we produce an ensemble of $1500$ IS's. We study five
different $T$ jumps,
with sets of $(T_{\rm i}\rightarrow T_{\rm b})$ including
$(0.55 \rightarrow 0.466)$,
$(0.6\rightarrow 0.1)$,
$(0.8\rightarrow 0.25)$,
$(0.8\rightarrow 0.446)$
and
$(0.8\rightarrow 0.6)$.

\begin{figure}
\centerline{
\includegraphics[width=2.8in]{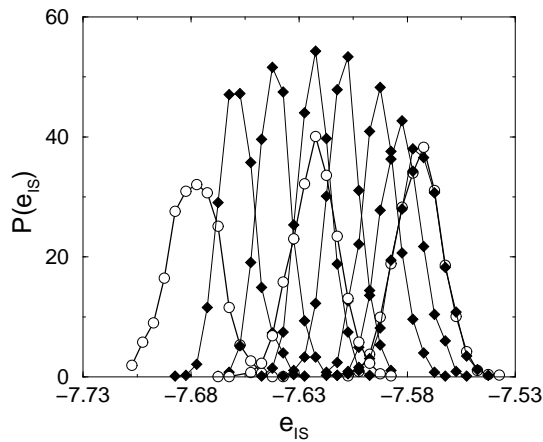}
}
\caption{Distributions of $e_{\rm IS}$ during aging.  Histograms (of unit area)
showing $P(e_{\rm IS},t_{\rm w})$ 
(filled diamonds) for an ensemble of aging systems with
$T_{\rm i}=0.8$ and $T_{\rm b}=0.446$ with
$t_{\rm w}= 542$, 3405, 7482, 21372, 46954, 174338 and 841500 
from right to left.  Also shown are equilibrium distributions 
$P(e_{\rm IS},T)$ (open circles) for $T=0.8$, 0.55 and 0.446
from right to left. }
\label{fig1}
\end{figure}

\begin{figure}
\centerline {
\includegraphics[width=2.8in]{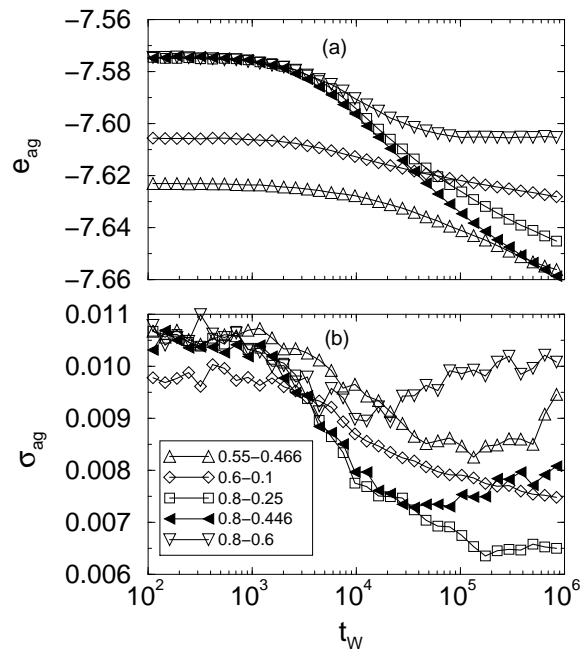}
}
\caption{Time evolution of the average $e_{\rm ag}$ (a)
and of the standard deviation $\sigma_{\rm ag}$ (b) of the 
inherent structure energy distribution 
$P(e_{\rm IS},t_{\rm w})$.
Legend indicates the corresponding $T_{\rm i}$ and $T_{\rm b}$ values.
}
\label{fig2}
\end{figure}

In Fig.~\ref{fig1} we plot $P(e_{\rm IS},t_{\rm w})$ for the 
$(0.8\rightarrow 0.446)$ run, as well as the
corresponding equilibrium distributions   $P(e_{\rm IS},T)$ at  
$T_{\rm i}$ and $T_{\rm b}$.  
The figure shows that distributions
initially become narrower (equivalently taller), then broader again 
on approaching equilibrium.
Fig.~\ref{fig2} shows the  average $e_{\rm ag}$ and the standard 
deviation $\sigma_{\rm ag}$  of the distributions 
as functions of $t_{\rm w}$.
In the cases where $T_{\rm b} > T_{\rm x}$, the variance,
which by construction at $t_{\rm w}=0$ coincides with the equilibrium 
variance, 
first decreases and then relaxes towards the equilibrium value.
For $T_{\rm b}< T_{\rm x}$, within the studied time window,
we only observe the narrowing. 
We also evaluate the skewness, though we can not discern any trend outside
of statistical noise.

Next we attempt to compare the fluctuations observed in aging with 
theoretical predictions. In equilibrium, the Helmholtz free energy per 
particle of the  liquid can be written as~\cite{sti82,sci99},
\begin{equation}\label{free-eq}
F = e_{\rm IS} -T S_c(e_{\rm IS}) + f_{\rm vib}(e_{\rm IS},T),
\end{equation}
where $f_{\rm vib}(e_{\rm IS},T)$ is the free energy of a basin, 
averaged over all  basins of depth  $e_{\rm IS}$, and
$S_{\rm c}(e_{\rm IS})=N^{-1} k_{\rm B} \ln \Omega (e_{\rm IS})$ is the 
configurational entropy.
$\Omega (e_{\rm IS})$ counts the number of basins of energy  $e_{\rm IS}$.
In the Gaussian-harmonic approximation (GHA)~\cite{nav03}, one assumes 
that $S_{\rm c}$ and $f_{vib}$ are quadratic in $e_{IS}$. 
In the GHA, $P(e_{\rm IS},T)$ is Gaussian with a 
$T$-independent variance, i.e.,
\begin{equation}\label{peist1}
P(e_{\rm IS},T) = \frac{1}{\sqrt{2\pi\sigma_{\rm P}^2}}
                  \exp{\left[-\frac{(e_{\rm IS} - 
                  \bar{e}_{\rm IS}(T))^2}{2\sigma_{\rm P}^2}\right]},
\end{equation}
where the $T$ dependence of $\bar{e}_{\rm IS}$
and the variance $\sigma_{\rm P}^2$ are expressed in terms of the 
statistical properties of the PES (the variance $\sigma^2_{\rm c}$ and 
the average energy $E_0$ of $\Omega(e_{\rm IS})$) 
and  of the parameters connecting depth and shape of the basins
($b$ and $c$)~\cite{notaM,params}. More precisely, $\bar{e}_{\rm IS}(T)= 
(E_0 - b\sigma_{\rm c}^2)/(1+2c\sigma_{\rm c}^2) 
     - T^{-1}\,\sigma_{\rm c}^2/(1+2c\sigma_{\rm c}^2)$, and
$\sigma_{\rm P}^2 = N^{-1}\sigma_{\rm c}^2/(1+2c\sigma_{\rm c}^2)$. 
Below $T=0.6$, the GHA provides a good description of the numerical  
data~\cite{sci99,sri01}, enabling us to estimate 
$\sigma^2_{\rm c}$, $E_0$, $b$ and $c$.

\begin{figure}
\centerline {
\includegraphics[width=2.8in]{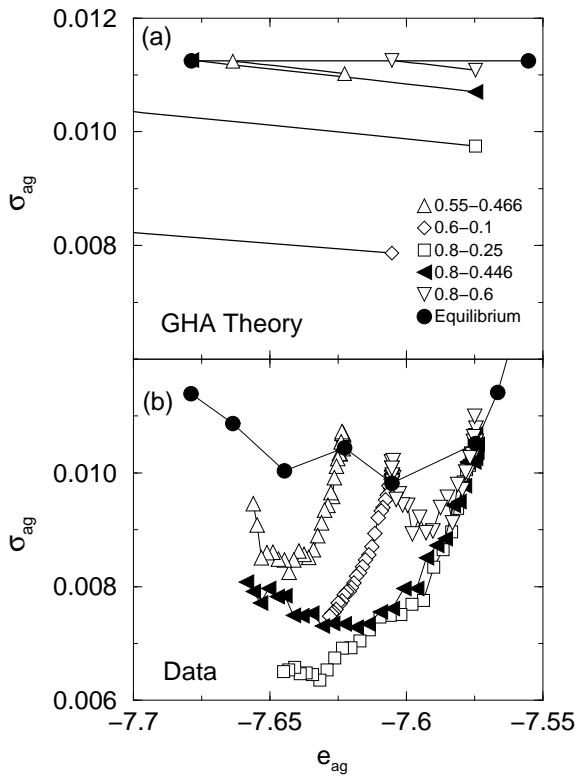}
}
\caption{Relationship between $\sigma_{\rm ag}$  and $e_{\rm ag}$ during 
aging.
The legend indicates $T_{\rm i}$ and $T_{\rm b}$. 
Panel (a) shows the theoretical
predictions based on the GHA, using the landscape parameters 
calculated fitting equilibrium data~\cite{params}. 
Panel (b) shows MD data.
Filled circles indicate equilibrium values.}
\label{fig3}
\end{figure}

In the non-equilibrium case, recent theoretical approaches suggest that
the free energy can be written as,
\begin{equation}\label{free}
F= e_{\rm IS} - T_{\rm f} S_{\rm c}(e_{\rm IS}) 
              + f_{\rm vib}(e_{\rm IS},T_{\rm b}),
\end{equation}
\noindent
where $S_{\rm c}$ is weighted by 
$T_{\rm f}$~\cite{fra00,fsfdt,lanavepre,verrocchio}.
The introduction of $T_{\rm f}$ modifies Eq.~\ref{peist1} 
to yield a non-equilibrium  probability density, in the GHA,
\begin{equation}\label{peistf}
P(e_{\rm IS},T_{\rm f}(t_w),T_{\rm b})
 = \frac{1}{\sqrt{2\pi\sigma_{\rm ag}^2}} 
 \exp{\left[ -\frac{(e_{\rm IS} - e_{\rm ag})^2}
 {2\sigma_{\rm ag}^2}\right]},
\end{equation}
where 
$e_{\rm ag}$ and 
$\sigma_{\rm ag}$ are given by
\begin{equation}
e_{\rm ag}=\frac{E_0 T_{\rm f}-\sigma_{\rm c}^2 - \sigma_{\rm c}^2 T_{\rm b}b}
                       {2c\sigma_{\rm c}^2 T_{\rm b}+ T_{\rm f}},
\label{teffGH}                   
\end{equation}
and
\begin{equation}\label{sigag}
\sigma_{\rm ag}^2=
N^{-1}\left(2c + \frac{1}{\sigma_{\rm c}^2} 
  \frac{T_{\rm f}}{T_{\rm b}}\right)^{-1}.
\end{equation}
Note that the theory predicts a gradual increase of the variance   
on increasing $t_{\rm w}$, i.e., on approaching equilibrium 
(decreasing $T_{\rm f}$).

The theoretical predictions for the relation 
between $e_{\rm ag}$  and $\sigma_{\rm ag}$, parametric in $T_{\rm f}$, 
are shown in Fig.~\ref{fig3}(a).
In Fig.~\ref{fig3}(b) we show the evolution of the same quantities as 
calculated from the aging runs. 
In the simulation data, at short $t_{\rm w}$, the  system  possesses by 
construction a distribution of IS energies identical to that of the 
starting $T$, and hence it is not surprising that the numerical results  
initially differ 
from the value predicted  by Eq.~\ref{sigag}.   
For longer times, i.e., for lower $e_{\rm IS}$ values, 
a region where the variance increases  is indeed observed, 
but only  for quenches above $T_{\rm x}$. 
The magnitude of change in the variance predicted by the theory  
is significantly  smaller than the one observed in simulation.   
Even in the case  $(0.55 \rightarrow 0.466)$, where the starting $T$ 
is such that the GHA should hold the best, the observed 
narrowing falls outside the uncertainty arising from modeling the 
equilibrium distributions,  as well as from numerical errors. 

\begin{figure}
\centerline {
\includegraphics[width=2.8in]{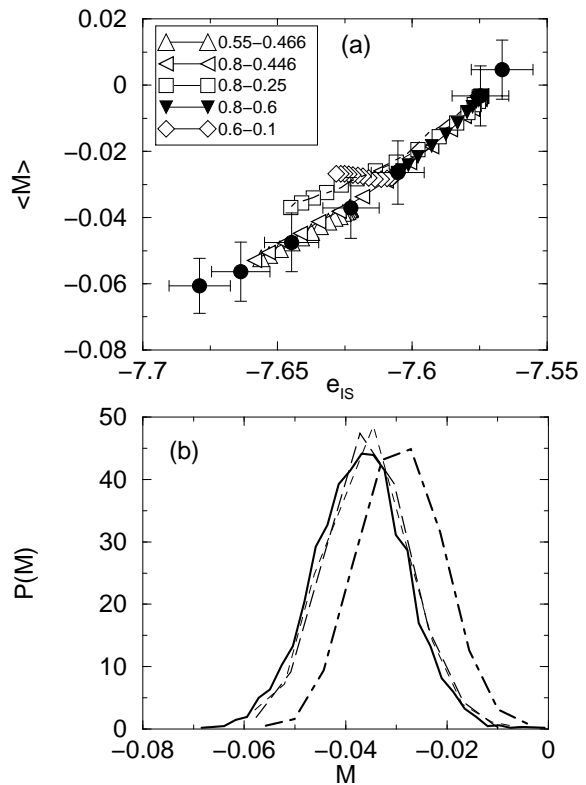}
}
\caption{Vibrational properties of basins as a function of $e_{\rm IS}$.
Panel (a) shows $\left< M\right>_{\rm eq}(e_{\rm IS})$ (filled circles,
bars indicating standard deviations) and
aging ensemble averages $\left< M \right>_{\rm ag}(e_{\rm IS})$.
For the dot-dash line we use a binning procedure to
determine $M(e_{\rm IS})$ from all basins regardless of $t_{\rm w}$
during aging for the $(0.8 \rightarrow 0.25)$ run.
Panel (b) shows the equilibrium $P(M)$ for $T=0.55$ (solid line),
and for the case where all basins with energy
$\bar{e}_{\rm IS}(T=0.55)\pm 0.002$ are used (short dash);
and for the aging runs 
$(0.8 \rightarrow 0.446)$ (long dash) and
$(0.8 \rightarrow 0.25)$ (dot dash), also binning the energy.
}
\label{fig4}
\end{figure}

An implicit assumption in the theory  is that, during aging, the 
liquid explores configurations typical of  equilibrium~\cite{kob00} 
and hence that the landscape parameters 
derived from the equilibrium study can be used  to
calculate the out-of-equilibrium distributions. Previous work has shown 
that this is not necessarily so~\cite{mossaB}. Here we analyze our 
extremely large database  of inherent structures in equilibrium and in aging
to  address the question of whether the basin curvature 
during aging differs from the equilibrium case. The quantity
 $M \equiv N^{-1} \sum_{i=1}^{3N-3} \ln{\frac{ \hbar \omega_i}
 {\epsilon_{\rm AA}}}$ 
constitutes a good indicator of the curvature of a single basin in the 
harmonic approximation~\cite{kob00,mossaB}.
In Fig.~\ref{fig4}(a)  we plot average values 
$\left<M\right>_{\rm eq}(e_{\rm IS})$ and 
$\left<M\right>_{\rm ag}(e_{\rm IS})$ for equilibrium and aging runs. 
This plot shows that for all aging experiments with $T_{\rm b}> T_{\rm x}$,  
the $e_{\rm IS}$ dependence of
$M$ is the same in equilibrium and in aging, and hence that 
the aging liquid explores basins the shape of which is the same as the 
one explored in equilibrium.  
In contrast, for $T_{\rm b}<T_{\rm x}$, the aging system  
explores basins of different shape. 
In the $(0.8 \rightarrow 0.25)$ case, the system begins to explore 
basins of higher $M$ 
relative to equilibrium  
values near $e_{\rm IS}\approx -7.6$. 
In the deep quench $({0.6}\rightarrow {0.1})$ case, 
the system stays in basins of roughly constant $f_{\rm vib}$. It is as 
though the system explores basins of lower energy, but not of correspondingly 
lower $f_{\rm vib}$.  
To discern the origins of the differences in basin sampling in 
aging and equilibrium, we calculate 
the distribution of $M$,  $P(M)$, for four different cases 
(Fig.~\ref{fig4}(b)).
More specifically, we want to assess if differences in $\left<M\right>$, 
at the same average $e_{\rm IS}$ value, arise from differences in the 
distributions of sampled $e_{\rm IS}$ or if they arise from intrinsic 
differences in sampled basin shapes. For this reason  we show, together 
with the equilibrium case for  $T=0.55$ 
(for which $\bar{e}_{\rm IS}=-7.6228$), the $P(M)$ for  basins 
of depth $e_{IS}=-7.6228 \pm 0.002$, 
explored  in the shallow $(0.8\rightarrow 0.446)$  and deep 
$(0.8\rightarrow 0.25)$  quenches 
(independent from $t_{\rm w}$) and basins explored in equilibrium  
(independent from $T$). 
Data in Fig.~\ref{fig4}(b) show that  the equilibrium $P(M)$ 
coincides with the  distribution evaluated from all sampled basins 
with depth equal to
$e_{IS}=-7.6228$.  It also coincides with the distributions
evaluated during aging in the $T_{\rm b}> T_{\rm x}$ runs.  
But it does not coincide with the distributions evaluated during aging in  
the $T_{\rm b}< T_{\rm x}$ case.
Hence, differences between equilibrium and aging can not be ascribed to 
the  the non-linear dependence of $M$ on $e_{\rm IS}$.
We conclude that  in aging runs with $T_{\rm b}< T_{\rm x}$ the system 
explores narrower 
basins (higher $M$) as compared to equilibrium.

In conclusion, the results reported in this Letter provide high quality 
data for the evolution of the energy distribution in aging,  
for several different aging scenarios,
and hence  provide an important starting
point for a detailed master equation 
description~\cite{dyre87,trap,recentkovacs,keyes}
of the dynamics in configuration space.
In our simulations, when $T_{\rm b} > T_{\rm x}$,
we find that the distribution  initially  narrows during aging, broadening 
again as the system re-equilibrates. We also find that the system samples 
configurations which are typically explored in equilibrium.
In this $T$ range, a  description of the aging dynamics via a master equation
is feasible, provided
equilibrium properties are fully characterized.
By contrast, when $T_{\rm b} < T_{\rm x}$, we only observe the narrowing of
the distribution.  Furthermore, the vibrational states explored during aging
differ from the equilibrium ones.
We note that the out-of-equilibrium theory,
which is built on the hypothesis that
states explored in aging are the same as those explored in equilibrium,
predicts only broadening associated with the approach to equilibrium.
Hence a theory based on a single $T_{\rm f}$ may only apply to the 
$T_{\rm b} > T_{\rm x}$ case.
We also stress that the results reported in this Letter 
suggest that  fluctuations in arrested states 
(which on the time scale of simulation correspond 
to $T_{\rm b}<T_{\rm x}$) depend on the previous thermal 
history~\cite{recentkovacs}. 
Glasses with the same average values may strongly differ in their frozen 
fluctuations. 
For deep quenches, the fluctuations frozen in the glass are significantly 
different from the 
fluctuations experienced in the liquid state, 
a feature which could explain the cross-over 
effect~\cite{mckenna89,mossa03,recentkovacs} and the failure of the 
theories based on one $T_{\rm f}$. 
Finally, we note that the decrease of the variance at short times 
(as shown in Fig.~\ref{fig2}),
suggests that the initial aging dynamics --- which may well be the only 
one accessible on the observation time scale for deep quenches ---
acts in the direction of increasing the differences between the 
equilibrium liquid and glass  distributions. 

We thank SHARCNET for computing resources.
I.S.-V. acknowledges NSERC for financial support.
We acknowledge support from MIUR Cofin 2002 and Firb
and INFM Pra GenFdt.


\begin{thebibliography}{999}

\bibitem{tool46} A.~Q. Tool,
J. Am. Ceram. Soc. {\bf 29}, 240 (1946).

\bibitem{brawer}
J. Brawer, {\it Relaxation in Viscous Liquids and
Glasses}
(Columbus, OH: The American Ceramics Society 1985).

\bibitem{scherer86}
G.~W. Scherer, {\em Relaxation in Glass and
Composites}
(Wiley, New York, 1986).

\bibitem{mckenna89}
G.~B. McKenna, in {\em Comprehensive Polymer
Science}, Vol.~2
Polymer Properties, edited by C.~Booth and C.~Price
(Pergamon, Oxford, 1989),
pp. 311-362.

\bibitem{cug97} L.~F. Cugliandolo {\it et al}.,
Phys. Rev. E {\bf 55}, 3898 (1997).


\bibitem{fra00} S. Franz and M.~A. Virasoro,
J. Phys. A {\bf 891}, 891 (2000).

\bibitem{nie01} Th.~M. Nieuwenhuizen,
Phys. Rev. Lett. {\bf 80}, 5580 (2001);
L. Leuzzi and Th.~M. Nieuwenhuizen,
Phys. Rev. E {\bf 64}, 011508 (2001).

\bibitem{fsfdt} F. Sciortino and P. Tartaglia,
Phys. Rev. Lett. {\bf 86}, 107 (2001).

\bibitem{mos02} S. Mossa {\it et al}.,
Eur. Phys. J. B {\bf 30}, 351 (2002); 
J. Phys. Condens. Matter {\bf 15}, S351 (2003).

\bibitem{sti82} F.~H. Stillinger and T.~A. Weber,
Phys. Rev. A {\bf 25}, 978 (1982);
Science {\bf 225}, 983 (1984);
F~H. Stillinger, Science {\bf 267}, 1935 (1995).

\bibitem{bacini} The set of particle configurations which map onto the
same local minimum via a path of steepest descent form a so-called basin.

\bibitem{sci99} F. Sciortino {\it et al}., 
Phys. Rev. Lett. {\bf 83}, 3214 (1999).

\bibitem{heu00} A. Heuer and S. B\"uchner,
J. Phys.: Condens. Matter {\bf 12}, 6535 (2000).

\bibitem{sri01} S. Sastry,
Nature {\bf 409}, 164 (2001).

\bibitem{kovacs}
A.~J. Kovacs,
Fortschr. Hochpolym. Forsch. {\bf 3}, 394 (1963).

\bibitem{mossa03}
S. Mossa and F. Sciortino,
cond-mat/0305526 (2003).

\bibitem{kob95} W. Kob and H.~C. Andersen,
Phys. Rev. E {\bf 51}, 4626 (1995); {\bf 52}, 4134 (1995).

\bibitem{units}
Here
$t_0=\sqrt(m\sigma_{\rm AA}/48\epsilon_{\rm AA})$, where $m$ is the particle
mass, and $\sigma_{\rm AA}$ and $\epsilon_{\rm AA}$, which enter in
the LJ potential, are the units of length and energy.

\bibitem{nav03} E. La Nave {\it et al}.,
J. Phys.: Condens. Matter {\bf 15}, S1085 (2003).

\bibitem{notaM} In the harmonic approximation,
 $f_{\rm vib}(e_{\rm IS},T)= k_{\rm B} T [\ln\left< e^{M}\right>
- 3(1-1/N)\ln(k_{\rm B}T/\epsilon_{\rm AA})]$, where
$M=N^{-1} \sum_{i=1}^{3N-3}
\ln{\left(\frac{ \hbar \omega_i}{\epsilon_{\rm AA}}\right)}$,
$N$ is the number of particles, and
$k_{\rm B}$ is the Boltzmann constant.

\bibitem{params}
By fitting $\bar{e}_{\rm IS}(T)$ and 
$\left<M\right>=\ln\left<e^M\right>=a+be_{\rm IS}+ce^2_{\rm IS}$,
we obtain
$a=166.34$,
$b=43.07$,
$c=2.787$,
$E_0 = -6.596$,
and
$\sigma_{\rm c}^2 = 0.43$.


\bibitem{lanavepre}
E. La Nave {\it et al}.,
Phys. Rev. E {\bf 68}, 032103 (2003).

\bibitem{verrocchio}
T.~S. Grigera {\it et al}.,
cond-mat/0307643 (2003).

\bibitem{kob00}
W. Kob {\it et al}.,
Europhys. Lett. {\bf 49}, 5906 (2000).

\bibitem{mossaB}
S. Mossa {\it et al}., 
Philos. Mag. B {\bf 82}, 695 (2002).

\bibitem{dyre87}
J.~C. Dyre,
Phys. Rev. Lett. {\bf 58}, 792 (1987);
Phys. Rev. B {\bf 51}, 12276 (1995).

\bibitem{trap} C. Monthus and J.-P. Bouchaud,
J. Phys. A: Math. Gen. {\bf 29}, 3847 (1996);
B.~Rinn {\it et al}.,
Phys. Rev. Lett. {\bf 84}, 5403 (2000).
R.~A. Denny {\it et al}.,
Phys. Rev. Lett. {\bf 90}, 025503 (2003).

\bibitem{recentkovacs}  E.~M. Bertin {\it et al}.,
 J. Phys. A: Math. Gen. {\bf 36}, 10701  (2003).

\bibitem{keyes} T.~Keyes {\it et al}.,
Phys. Rev. E {\bf 66}, 051110 (2002).




\end{thebibliography}
\end{document}